\documentstyle[12pt]{article}
\def\one{1\hskip-.37em 1}

\def\half{\textstyle{\frac{1}{2}}}
\def\quarter{\textstyle{\frac{1}{4}}}

\def\ra{\rightarrow}
\def\H{{\cal H}}
\def\tint{\textstyle\int}
\def\D{{\cal D}}

\def\o{\overline}
\def\b{\begin{eqnarray*}}     
\def\e{\end{eqnarray*}}       
\def\bn{\begin{eqnarray}}     
\def\en{\end{eqnarray}}       
\def\<{\langle}
\def\>{\rangle}

\def\{{\lbrace}
\def\}{\rbrace}

\title{Coherent States in Action\footnote{Based on a contribution to the 
International Conference on Frontiers in Quantum Physics, Kuala Lumpur, 
Malaysia, July, 1997.}}
\author{John R. Klauder\\
Departments of Physics and Mathematics\\
University of Florida\\
Gainesville, Fl  32611}
\date{}
\begin{document}
\maketitle
\begin{abstract}
Quantum mechanical phase space path integrals are re-examined with regard to 
the physical interpretation of the phase space variables involved. It is 
demonstrated that the traditional phase space path integral implies a meaning 
for the variables involved that is manifestly inconsistent. On the other 
hand, a phase space path integral based on coherent states entails variables 
that exhibit a self-consistent physical meaning.
\end{abstract}
\section{Conventional phase space path integrals}
There is considerable appeal in the formal phase space path integral
 \bn\<q''|\,e^{-i\H T }\,|q'\>={\cal N}\int e^{i\tint[p{\dot q}-h(p,q)]\,dt}
\,\D p\,\D q  \en
which yields the propagator in the $q$-representation \cite{fey}. In this 
relation the integration is over all $q$-paths $q(t)$, $t'\leq t\leq t''
\equiv t'+T$, $T>0$, subject to the boundary conditions that $q(t'')=q''$ 
and $q(t')=q'$, as well as all $p$-paths $p(t)$ for $t'\leq t\leq t''$. It 
follows from this formula that the {\it meaning} of $q(t)$ is the same as 
the meaning of $q(t'')$, namely, as the sharp eigenvalue of the position 
operator $Q$, where $Q|q\>=q|q\>$.

An analogous path integral leads to the propagator in the $p$-representation 
and is given by
 \bn \<p''|\,e^{-i\H T }\,|p'\>={\cal N}\int e^{i\tint[-q{\dot p}-h(p,q)]
\,dt}\,\D p\,\D q\;.  \en
In this expression integration runs over the $p$-paths $p(t)$, $t'\leq t
\leq t''$, subject to the requirement that $p(t'')=p''$ and $p(t')=p'$, 
while in the present case, integration over all $q$-paths $q(t)$, $t'\leq t
\leq t''$, is assumed. It follows that the {\it meaning} of $p(t)$ is the 
same as the meaning of $p(t'')$, namely as the sharp eigenvalue of the 
momentum operator $P$, where $P|p\>=p|p\>$.

These two path integrals are of course connected with each other. In 
particular, it follows that
\bn&&\hskip-2.5cm\<q''|\,e^{-i\H T}\,|q'\>
 ={(2\pi)}^{-1}\int e^{i(q''p''-q'p')}\,\<p''|\,e^{-i\H T}\,|p'\>\,dp''
\,dp'\nonumber\\
&&={\cal N}\int e^{i\tint[{\dot{\o {pq}}}-q{\dot p}-h(p,q)]\,dt}\,\D p\
,\D q\nonumber\\
  &&={\cal N}\int e^{i\tint[p{\dot q}-h(p,q)]\,dt}\,\D p\,\D q  \en
just as before. 

Is the so obtained physical meaning for $p(t)$ and $q(t)$ satisfactory? 
If we were dealing with the strictly classical theory, for which $\hbar=0$, 
there is absolutely no contradiction in specifying $p(t)$ and $q(t)$ 
simultaneously for all $t$, $t'\leq t\leq t''$. On the other hand, we are 
dealing with the quantum theory and decidedly not the classical theory. 
Planck's constant $\hbar=1$ (in the chosen units) and does not vanish. T
hus we are led to the conclusion that the given formal path integrals are 
expressed in terms of phase space paths for which, {\it within the quantum 
theory}, one may simultaneously specify both position $q(t)$ and momentum 
$p(t)$, $t'< t< t''$ sharply. This assertion evidently contradicts the 
Heisenberg uncertainty principle, and consequently it is unacceptable. 
Something is definitely wrong!

Another indication that something is wrong follows on consideration of the 
expression
  \bn{\cal N}\int e^{i\tint[\frac{1}{2}(p{\dot q}-q{\dot p})-h(p,q)]\,dt}
\,\D p\,\D q\;, \en
which also involves an acceptable version of the classical action, but which 
cannot be interpreted along the lines given above. Interpretation fails 
because it is unclear what variable(s) are to be held fixed at the initial 
and final times. For instance, should this expression be interpreted as
 \bn C\int e^{i(p''q''-p'q')/2}\,\<p''|\,e^{-i\H T}\,|p'\>\,dp''\,dp'\;,\en
where $C$ is an appropriate constant, or as 
\bn C\int e^{-i(p''q''-p'q')/2}\,\<q''|\,e^{-i\H T}\,|q'\>\,dq''\,dq'\en
either of which would seem to be equally possible interpretations but which 
evidently lead to unequal results. 
\subsection{Why do interpretational problems exist?}
The reason these expressions lead to inconsistencies of interpretation is 
really very simple---although it is a reason that physicists are often 
reluctant to entertain. The argument presented above fails because the 
{\it indicated representations} for $\<q''|e^{-i\H T}|q'\>$ and $\<p''|
e^{-i\H T}|p'\>$ simply {\it do not exist} as given. Physicists tend to 
believe that if they can write down a set of relations possessing a {\it 
formal} self consistency, then the underlying existence of the relations is 
not in doubt.\footnote{A simple but informative example of this issue is the 
following. Let $\{1,2,3,...\}$ denote the set of positive integers. Let $X$ 
denote the largest such integer, and let us assume that $X>1$. Since $X^2>X$, 
we observe there is an integer larger than $X$, therefore we conclude that 
our assumption that $X>1$ was in error, hence $X=1$.} Of course, the dilemma 
surrounding these relations can be lifted by giving alternative 
representations that, in fact, do exist. One such representation is based 
on a lattice limit, namely, by giving meaning to the undefined formal path 
integral as the limit of a sequence of well defined finite dimensional 
integrals. As one such prescription we offer \cite{sch,roe}
 \bn &&\hskip-.4cm\<q''|\,e^{-i\H T}\,|q'\>\nonumber\\
&&=\hskip-.1cm\lim_{N\ra\infty}\frac{1}{(2\pi)^{N+1}}\int\exp\{i
\Sigma_{l=0}^N[ p_{l+1/2}(q_{l+1}-q_l)-\epsilon h(p_{l+1/2},(q_{l+1}+q_l)/2)]
\}\nonumber\\
&&\hskip4cm\times\Pi_{l=0}^N\,dp_{l+1/2}\,\Pi_{l=1}^N\,dq_l\;. \en
Here the limit $N\ra\infty$ also implies that $\epsilon\equiv T/(N+1)\ra0$, 
and $q_{N+1}$ $\equiv q''$ and $q_0\equiv q'$; all $p$ values are integrated 
out.
This prescription, which applies for a wide class of classical Hamiltonian 
functions $h(p,q)$, generates the propagator in the Schr\"odinger 
$q$-representation, and two such propagators properly fold to a third 
propagator when integrated over the intermediate $q$ with a measure $dq$. 

However, this is not the only prescription that can be offered for the same 
formal phase space path integral. 
\section{Coherent state formulation}
Another rule of definition that can also be accepted for the formal phase 
space path integral (1) is given by \cite{kla1}
 \bn&&\hskip-.5cm\<p'',q''|\,e^{-i\H T}\,|p',q'\>\nonumber\\
&&\equiv\lim_{N\ra\infty}\frac{1}{(2\pi)^N}\int\exp(\!\!(\Sigma_{l=0}^N\{i
\half(p_{l+1}+p_l)(q_{l+1}-q_l)\nonumber\\
&&\hskip2cm-\quarter[(p_{l+1}-p_l)^2+(q_{l+1}-q_l)^2]\nonumber\\
&&\hskip2cm-i\epsilon h(\half(p_{l+1}+p_l+iq_{l+1}-iq_l),\half(q_{l+1}+q_l
- -ip_{l+1}+ip_l))\})\!\!)\nonumber\\
&&\hskip3cm\times\Pi_{l=1}^N\,dp_l\,dq_l\;.   \en
This expression differs from the former one in that both $p$ {\it and} $q$ 
are held fixed at the initial {\it and} final end points. In particular, 
now $(p_{N+1},q_{N+1})$ $\equiv (p'',q'')$ and $(p_0,q_0)\equiv(p',q')$. Two 
such propagators properly fold together to a third propagator with an 
integration over the intermediate variables $p$ and $q$ with the measure 
$dp\,dq/2\pi$. Like the previous case, the present expression holds for a 
wide class of classical Hamiltonian functions. However, this latter sequence 
is {\it fundamentally} different than the previous one, and that difference 
not only involves a different sort of representation but even goes so far as 
to entail a profound change of the {\it meaning} of the symbols $p$ and $q$ 
from their meaning as found in the preceding section. 

The states $|p,q\>$ implicitly introduced above are {\it canonical coherent 
states} defined by the following expression
  \bn |p,q\>\equiv e^{-iqP}e^{ipQ}\,|0\>\;,  \en
where, as usual, $[Q,P]=i\one$ and $|0\>$ denotes the ground state of a 
harmonic oscillator, i.e., a normalized solution of the equation $(Q+iP)|0\>
=0$ \cite{kla2}. Observe in this case that neither $p$ nor $q$ are {\it 
eigenvalues} of any operator. Instead, it follows that
  \bn\<p,q|P|p,q\>=p\;,\hskip2cm\<p,q|Q|p,q\>=q\;,  \en
namely, that the labels $p$ and $q$ have the meaning of {\it expectation 
values} rather than eigenvalues. Thus there is absolutely no contradiction 
with the Heisenberg uncertainty principle in specifying both $p$ and $q$ 
{\it simultaneously}. The overlap of two coherent states, given by
\bn \<p',q'|p,q\>=\exp\{i\half(p'+p)(q'-q)-\quarter[(p'-p)^2+(q'-q)^2]\}\;,
\en
serves as a reproducing kernel for the functional Hilbert space 
representation in the present case. The folding of two such overlap 
functions leads to
\bn \int\<p'',q''|p,q\>\<p,q|p',q'\>\,dp\,dq/2\pi=\<p'',q''|p',q'\>\;, 
\en
an expression which shows that the coherent state overlap function serves as 
the ``$\delta$-function'' in the present representation although, of course, 
in the present case it is a bounded, continuous function. In short, we learn 
that the choice of sequential definition adopted to give meaning to the 
formal phase space path integral can lead to a dramatic change of 
representation and even of the meaning of the variables involved. 

We conclude these remarks with the observation that if we formally 
interchange the limit and integrations in (8) and write for the integrand 
the form it assumes for continuous and differential paths, the
result has the formal expression (1), namely
  \bn {\cal N}\int e^{i\tint[p{\dot q}-h(p,q)]\,dt}\,\D p\,\D q\;, \en
which is just the expression we started with! It is in this sense that
we assert that the present sequential definition is just as valid as the
one customarily chosen. Moreover, with the present understanding of the 
sequential definition, there is absolutely no conflict between the meaning 
of the variables $p$ and $q$ and the Heisenberg uncertainty principle; in 
the present case, $p$ and $q$ denote expectation values in the coherent 
states involved, and these may both be specified as general functions of 
time $p(t)$ and $q(t)$, $t'\leq t\leq t''$. 
 
It is clear to this author---but apparently unclear to many others---that 
the interpretation of the formal path integral (13) in terms of paths $p(t)$ 
and $q(t)$ for which the meaning of the variables is that of {\it expectation 
values} is far more acceptable than the one in which the meaning is that of 
both sharp {\it position} and sharp {\it momentum} (eigen)values. Even if 
one carries to the continuum the insight gained on the lattice for the usual 
formulation, namely, that $p$ and $q$ are  diagonalized {\it alternately} on 
successive time slices, the result is that the continuum interpretation is 
strictly not one for which $p$ and $q$ are {\it simultaneously} sharp but 
one where $p$ and $q$ are alternately sharp at every ``other'' instant of 
time---and of course when $p(q)$ is sharp then $q(p)$ is completely unknown! 
This is the true physical meaning of the variables entering the putative 
formal phase space path integral with the usual interpretation. How bizarre 
that interpretation is when it is fully appreciated for what it is!

Contrast the interpretation just outlined with the one appropriate to the 
alternative scenario in terms of canonical coherent states. In the case of 
a lattice formulation of the phase space path integral in terms of coherent 
states, $p$ and $q$ are specified at each time slice simultaneously and 
interpreted as expectation values. This interpretation persists in the 
continuum limit, and there is no logical conflict of that interpretation 
in such a limit. Moreover, there is a symmetry in the interpretation and 
usage of $p$ and $q$ inherent in the coherent state formulation that is 
simply unavailable in the more traditional formulation.

One is almost tempted to assert that the usual
interpretation in terms of sharp eigenvalues is ``wrong'', because it cannot 
be consistently maintained, while the interpretation in terms of expectation 
values is ``right'', because it can be consistently maintained.
On the other hand, the community at large may not be ready to swallow such a 
heretical statement, so perhaps it would be best if it was stricken from the 
record! However, before completely striking it from the record, it may not be 
inappropriate to offer additional evidence as food for thought.
\section{Wiener measure regularization}
We have accepted the fact that (13) is without mathematical meaning as it 
stands. Some sort of regularization and removal of that regularization is 
needed to give it meaning. There are many ways to do so, two of which have 
been illustrated above. In this section we discuss quite a different form of 
regularization.

Consider the expression \cite{daub}
\bn \lim_{\nu\ra\infty}{\cal N}\int\exp\{i\tint[p{\dot q}-h(p,q)]\,dt\}
\,\exp[-(1/2\nu)\tint({\dot p}^2+{\dot q}^2)\,dt]\,\D p\,\D 
q\;. \en
This expression differs from the usual one (13) by the presence of a damping 
factor---a convergence factor---involving the time derivative of both $p$ 
and $q$. The result of interest is given in the limit that the parameter 
$\nu\ra\infty$. Note that when $\nu=\infty$, formally speaking, the usual 
formal path integral (13) is recovered. Although (14) is written in the same 
formal 
language as (13), the latter expression is in fact profoundly different. In 
fact, (14) is intended to be a {\it regularized} form of (13). Admittedly, it 
doesn't appear any better defined than the usual expression in its present 
form; however, (14) can be given an alternative but equivalent formulation 
when we group certain terms together. In particular, with a suitable 
regrouping of terms (14) becomes
 \bn \lim_{\nu\ra\infty}(2\pi)\,e^{\nu T/2}\int e^{i\tint[p\,dq-h(p,q)\,dt]}
\,d\mu^\nu_W(p,q)\;.  \en
In this expression $\mu^\nu_W$ denotes (pinned) Wiener measure on the two 
dimensional plane expressed in Cartesian coordinates $(p,q)$. In addition, 
$p(t)$ and $q(t)$, $t'\leq t\leq t''$, denote Brownian motion paths with 
$\nu$ as the diffusion constant, and $\tint p\,dq$ denotes a stochastic 
integral needed since although $p(t)$ and $q(t)$ are continuous functions 
for all $\nu$ they are nowhere differentiable. 
For convenience we adopt the Stratonovich (midpoint) definition of the 
stochastic integral (which is equivalent to the It\^o definition in the 
present case because $dp(t)dq(t)=0$ is a valid It\^o rule in these 
coordinates). With those remarks the integral in (15) is a well defined 
expression for each $\nu$ and one may ask the question whether the 
indicated limit converges and if so whether that limit has anything to do 
with a solution to the Schr\"odinger equation. For a dense set of 
Hamiltonians the answer to both of these questions is {\it yes!}

However, before we relate this expression to the earlier discussion let us 
take up the possible meaning of the variables $p$ and $q$ as they appear 
in (15). Observe, as noted, that the expression is well defined as 
it stands---indeed, it involves a {\it continuous time regularization}. Thus 
if this expression is going to have something to do with quantum mechanics it 
must be consistent to simultaneously specify both $p(t)$ and $q(t)$ for all 
$t$ in the appropriate interval. This means that $p$ and $q$ {\it cannot} 
have the meaning of sharp momentum and sharp position, respectively. On the 
other hand, it would be possible for those variables to have the meaning of 
expectation values which can be simultaneously given. It should thus come as 
not too great a surprise that the continuous time regularization of a phase 
space path integral with the help of a Wiener measure on the plane, in the 
limit as the diffusion constant diverges, {\it automatically generates a 
coherent state representation!} 

In particular, with the Brownian paths pinned so that $p(t'')=p''$, $q(t'')
=q''$ and $p(t')=p'$, $q(t')=q'$, the resultant limit is equivalent to
\bn &&\hskip-1cm\<p'',q''|\,e^{-i\H T}\,|p',q'\>\nonumber\\
 &&=\lim_{\nu\ra\infty}(2\pi)\,e^{\nu T/2}\int e^{i\tint[p\,dq-h(p,q)\,dt]}
\,
d\mu_W(p,q)\;,  \en
where, as implied by (16) itself, and consistent with the earlier notation,
  \bn  &&|p,q\>\equiv e^{-iqP}e^{ipQ}|0\>\;,\hskip1cm(Q+iP)|0\>=0\;,\\
  &&\hskip.9cm\H\equiv\tint h(p,q)|p,q\>\<p,q|\,dp\,dq/2\pi \;.  \en
In other words, the result of the Wiener measure regularized phase space 
path integral, in the limit that the diffusion constant diverges, yields a 
propagator in the coherent state representation as we had discussed earlier. 
Here is an additional argument for favoring the interpretation of the formal 
phase space path integral as really standing for the one expressed in terms 
of coherent states rather than one that is internally inconsistent, namely, 
one interpreted in terms of sharp eigenvalues for the position and momentum. 

If one accepts the idea that the formal expression (13) may be best 
interpreted in terms of coherent states rather than sharp Schr\"odinger 
eigenstates, one may be worried that many previous calculations are 
incorrect. There is no need to worry. All previous calculations which are 
implicitly consistent with a lattice limit such as in (8) are perfectly 
correct. Our discussion is not addressed to revising the evaluation of 
properly interpreted path integrals but rather to stressing the 
consistency---or possible inconsistency---of interpreting the continuum 
version of the phase space path integral. With the coherent state 
interpretation one is completely justified in regarding the paths as 
functions defined for a continuous time parameter, and indeed within the 
sequence where $\nu<\infty$, as continuous functions of time. This is a 
{\it conceptual} difference with respect to how the interpretation in the 
usual formulation can be taken. If there is ever any hope to define path 
integrals rigorously as path integrals over a set of paths (functions of 
time), then it is {\it essential} to give up the notion that the paths 
involved are sharp value paths and replace that with another interpretation 
of which the expectation value paths is a completely satisfactory example. 
In point of fact, the present author feels that the rigorous definition (16) 
in terms of a limit of a sequence of {\it completely unambiguous} path 
integrals is as close as one is likely to come to a rigorous definition of a 
continuous time path integral in terms of genuine (i.e., countably additive) 
measures. One can hardly ask for an expression without some sort of 
regularization. For example, even the one dimensional integral
  \bn \int_{-\infty}^\infty e^{iy^2}\,dy  \en
is effectively undefined since it is only conditionally convergent. It, too, 
needs
a rule to overcome this ambiguity, and one rule is to define it as
\bn  \lim_{\nu\ra\infty}\int_{-\infty}^\infty e^{iy^2-y^2/\nu}\,dy\;. \en
The indicated sequence exists and the limit converges, but it has required 
the use of a convergence factor; one could hardly expect a real time path 
integral to require anything less!
\subsection{Generalization to non-flat phase spaces}
The point we are making here naturally leads to another line of thought on 
which we shall comment but not develop since it has been adequately treated 
elsewhere. If we are dealing with a conditionally convergent integral, then 
it is possible to obtain fundamentally different answers by choosing a 
qualitatively different form of regularization. In particular, from the 
point of view of regularization, why was it necessary for us to choose a 
Brownian motion regularization on a phase space that constitutes a {\it flat} 
two-dimensional space; why not consider a Brownian motion regularization on 
a phase space that is a curved two-dimensional manifold, say, a sphere or a 
pseudo-sphere, for example, or even a space of non-constant curvature. 
Brownian motion regularization on such non-flat spaces has indeed been 
investigated, and the result is most interesting. For a sphere (of the right 
radius) the result of the limit of the regulated phase space path integrals 
over continuous paths leads to a quantization in which the kinematical 
variables are spin variables, i.e., operators that obey the commutation 
relations of the Lie algebra of the group SU(2) \cite{daub}. If a 
pseudo-sphere is used instead, the result for the kinematical variables is 
that for the Lie algebra of the group SU(1,1) (or the ``$ax+b$'' group) 
\cite{daub2,kla3}. Both of these cases lead to group related coherent states 
and a representation of the propagator in terms of those states. On the other 
hand, for Brownian motion on a space without any special symmetry, the result 
again leads to coherent states \cite{ali}, but these are coherent states of 
a more general kind than traditionally considered since they are {\it not} 
associated with any group!

The moral of this extended story is that phase space path integrals of an 
exceedingly general kind appropriate to very general kinematical variables 
can be rigorously developed with the aid of a Weiner measure regularization 
each of which involves coherent states wherein the variables are {\it never} 
eigenvalues of some self-adjoint operator but more typically are associated 
with expectation values of suitable operators for which there is no 
conceptual difficulty in their simultaneous specification. This very 
desirable state of affairs has arisen by combining the {\it symplectic} 
geometry of the classical theory with a {\it Riemannian} geometry needed to 
carry the {\it Brownian motion} that forms the regularization. 

If we may be allowed a single phrase of summary, then it is no exaggeration 
to claim \cite{kla5} that, when properly interpreted, \vskip.3cm
\hskip1cm QUANTIZATION$\;=\:$GEOMETRY$\;+\;$PROBABILITY
 \section{Acknowledgments}
It is a pleasure to thank the organizers of the conference in Kuala Lumpur, 
especially Prof. S.C. Lim, for their fine organization and the splendid 
meeting that resulted.

\end{document}